\documentclass[]{iopart}
\usepackage{iopams}
\usepackage{amssymb}
\usepackage{graphicx}
\usepackage{psfig}
\usepackage{portland}
\def\ltord{\hbox{$\;\raise.4ex\hbox{$<$}\kern-.75em\lower.7ex\hbox{$\sim$}
                       \;$}}
\def\gtord{\hbox{$\;\raise.4ex\hbox{$>$}\kern-.75em\lower.7ex\hbox{$\sim$}
                       \;$}}

\begin{document}
\title[Computations of primordial black hole
formation]{Computations of primordial black hole formation}
\author{Ilia Musco${}^{1}$, John C. Miller${}^{1,2}$ and 
Luciano Rezzolla${}^{1,3}$}

\address{${}^1$ SISSA, International School for Advanced Studies and INFN, 
Via Beirut 2-4, 34014 Trieste, Italy;\\
${}^2$ Department of Physics (Astrophysics), University of Oxford, Keble
Road, Oxford OX1 3RH, England;\\
${}^3$ Department of Physics, Louisiana State University, Baton Rouge, LA 
70803, USA.\\
}

\begin{abstract}
 Results are presented from general relativistic numerical computations of
primordial black-hole formation during the radiation-dominated era of the
universe. Growing-mode perturbations are specified within the linear
regime and their subsequent evolution is followed as they become
nonlinear. We use a spherically symmetric Lagrangian code and study both
super-critical perturbations, which go on to produce black holes, and
sub-critical perturbations, for which the overdensity eventually disperses
into the background medium. For super-critical perturbations, we confirm
the results of previous work concerning scaling-laws but note that the
threshold amplitude for a perturbation to lead to black-hole formation is
substantially reduced when the initial conditions are taken to represent
purely growing modes. For sub-critical cases, where an initial collapse is
followed by a subsequent re-expansion, strong compressions and
rarefactions are seen for perturbation amplitudes near to the threshold.
We have also investigated the effect of including a significant component
of vacuum energy and have calculated the resulting changes in the
threshold and in the slope of the scaling law.
 \end{abstract}

\today

\section{Introduction}
\label{intro}
 Cosmological structure formation is thought to have resulted from the
growth and evolution of small perturbations initiated at the time of
inflation (see \cite{Liddle2} and references therein). Inflationary models
give rise to a spectrum of fluctuations on scales larger than the
cosmological horizon which then start to re-enter the horizon in the
radiation-dominated era. Primordial black holes (PBHs) could be formed at
this stage in extreme cases where the fluctuation amplitude exceeds a
critical threshold value \cite{Zeldovich,Hawking,Carr1}. The masses of
these PBHs could, in principle, span many orders of magnitude, from the
Planck mass up to the horizon mass at the time of equivalence between
radiation and matter ($\sim 10^4$ years after the Big Bang).

Sufficiently small PBHs would have emitted a significant amount of Hawking
radiation and, according to the standard picture, would have evaporated
completely away before now if their mass was less than $\sim 10^{15}$ g.  
It has been suggested that extremely small PBHs of $\sim 10^5 - 10^6$ g,
formed in the very early Universe, might have been the cause of
baryogenesis \cite{Nagatani}, while PBHs of $\sim 10^{15}$ g evaporating
now might explain a class of observed very short-duration gamma ray bursts
\cite{Cline1,Cline2,Green2} and some cosmic rays \cite{Carr4, Barrau}. It
has also been suggested that the evaporation of small black holes might
not continue all the way to zero mass but might stop in the region of the
Planck mass, and that surviving PBH remnants might explain some or all of
the current cold dark matter \cite{McGibbon,Blais,Afshordi}.

Larger PBHs with masses well above $\sim 10^{15}$ g could be observable by
means of gravitational effects such as their contribution to the
cosmological density parameter \cite{Carr2,Kim1}, gravitational radiation
emitted from coalescing binary PBH systems \cite{Nakamura} or
micro-lensing (PBHs were suggested as possible candidates for some of the
objects observed by the MACHO Collaboration \cite{MACHO} and subsequent
surveys, although the idea that they might comprise a significant fraction
of the halo dark matter is now no longer favoured \cite{Green4,Rahvar}).  
All of these aspects have been studied and constraints obtained both on
the fraction of matter in the Universe that could now be in the form of
PBHs and also on the spectral index of fluctuations on small scales
\cite{Carr3,Green1,Liddle1,Kribs,Bringmann}.

In this article, we present results from general relativistic numerical
simulations of PBH formation in the background of an expanding universe in
the radiation-dominated era. There have been a number of previous studies
of this type and our aim here is to re-visit the subject area,
highlighting some features which we think are important and interesting. A
particular aspect of our study is that we use initial perturbations
representing growing-mode overdensities with length-scales larger than
the horizon and still within the linear regime. Their evolution is
then followed as they subsequently become nonlinear. A convenient
parameter for measuring perturbation amplitudes is the fractional
mass-excess within the overdense region (denoted by $\delta$); if this is
greater than a critical value $\delta_c$ (super-critical perturbations),
the following nonlinear evolution will result in formation of a black
hole, whereas if it is smaller than $\delta_c$ (sub-critical
perturbations), the overdensity will eventually disperse back into the
surrounding medium. Determining the value of $\delta_c$ is very important
for the cosmological considerations mentioned above.

Following the earliest papers on this subject \cite{Zeldovich,Hawking,
Carr1}, Carr (1975) \cite{Carr2} carried out a quantitative study based on
a simplified model consisting of an overdense collapsing region, described
by a closed Friedmann-Robertson-Walker (FRW) spacetime, surrounded by a
spatially-flat FRW expanding background. For a radiation-dominated
universe, his calculation gave $\delta_c = 1/3$ and the black holes formed
had masses $M_{BH}$ of the same order as the horizon mass at the time when
the fluctuation first entered the horizon. Nadezhin, Novikov \& Polnarev
(1978) \cite{Nadezhin} carried out the first detailed numerical study of
PBH formation using a hydrodynamical computer code similar to those of May
\& White (1966) \cite{May} and Podurets (1964) \cite{Podurets} using a
``cosmic-time'' coordinate with a diagonal metric which reduces to a form
similar to that of the FRW metric in the absence of perturbations. A
well-known difficulty with this approach is that in a continuing collapse,
singularities typically appear rather quickly and stop the computation
before the black hole formation is complete. In \cite{Nadezhin}, this
difficulty was overcome by using an early form of excision with the
evolution being stopped in the region where the singularity would appear.
The qualitative features of the earlier picture were basically confirmed
but it was found that the PBH masses were always much smaller than the
horizon mass.

Shortly afterwards, Bicknell \& Henriksen (1979) \cite{Bicknell} carried
out related calculations using a method based on integration along
hydrodynamical characteristics which avoids the problems associated with
the appearance of singularities. They used rather different initial data
from that of \cite{Nadezhin} and found formation of black holes with
masses of the same order as the horizon mass (or greater in cases where
the overdensity in the initial perturbation was not compensated by a
surrounding under-dense region). They noted the appearance of both ingoing
and outgoing compression waves during evolutions leading to black hole
formation. 

More recently, Niemeyer \& Jedamzik \cite{Niemeyer1,Niemeyer2} made
further numerical calculations, particularly focusing on the relevance of
scaling-laws for PBH formation. They found that $M_{BH}$ follows a power
law in $(\delta - \delta_c)$ when the latter is sufficiently small, which
is a similar behaviour to that seen in critical collapse by Choptuik
\cite{Choptuik} and many subsequent authors (see Evans \& Coleman
\cite{Evans} and the review by Gundlach \cite{Gundlach}). They started
from initial perturbations specified at the moment when the overdensity
enters the horizon and then computed the subsequent evolution. They used a
null-slicing code, following the formulation by Hernandez \& Misner (1966)
\cite{Hernandez}, and obtained $\delta_c \simeq 0.7$ for each of the three
types of perturbation profile which they studied.

While Niemeyer \& Jedamzik \cite{Niemeyer2} demonstrated the existence of
a scaling law for PBH masses down to around one tenth of the horizon mass,
they did not investigate smaller masses and so it was not possible to
determine whether the scaling law was likely to continue down to
vanishingly small masses (when $\delta \to \delta_c$) as in type II
critical collapse. In fact, the calculations become very challenging from
a numerical point of view when $\delta$ is very close to $\delta_c$
because of the appearance of strong shocks and deep voids outside the
region where the PBH is forming. Hawke \& Stewart (2002) \cite{Hawke}
addressed this problem using a sophisticated purpose-built code which
allowed them to make calculations for values of $\delta$ closer to
$\delta_c$ and to handle the strong shocks which occur in these cases.
They found that the scaling law does {\em not} continue down to very small
values of $(\delta - \delta_c)$ but rather reaches a minimum value for
$M_{BH}$ around $10^{-4}$ of the horizon mass, with the limit resulting
from the behaviour of the shocks produced in nearly critical collapse. 

Shibata \& Sasaki (1999) \cite{Shibata} presented an alternative formalism
for studying PBH formation using constant mean curvature time slicing and
focusing on metric perturbations rather than density perturbations. They
emphasised the importance of using initial data which can be directly
related to perturbations arising from inflation. Their formulation was not
restricted to spherical symmetry (as had been the case for the previous
authors mentioned here) but they presented results only from spherical
calculations. They located the threshold perturbation amplitude for PBH
formation (in terms of the metric perturbation) and found that this varied
considerably depending on the density of the medium surrounding the
density peak. They concluded from this that it is probably important to
take into account spatial correlations of density fluctuations when
considering PBH formation.

It is not straightforward to make the link between results from this type
of calculation and ones from the more standard approach focused on
density fluctuations. However, this was addressed in a recent paper by
Green et al. (2004) \cite{Green3}. They calculated the PBH abundance
produced from two different fluctuation spectra, using peaks theory
together with the threshold criterion of Shibata \& Sasaki \cite{Shibata}.  
They then compared the results of this with ones obtained from a standard
calculation based on a Press-Schechter-like approach and using density
perturbations. They found that the Shibata \& Sasaki results are
consistent with ones using a density perturbation if $\delta_c$ lies in
the range $0.3 \ltord \delta_c \ltord 0.5$ and they noted the discrepancy
with the results of \cite{Niemeyer2}.

As mentioned previously, our aim in the work described in this paper was
to re-visit some issues arising from these earlier calculations. In
particular, we wanted to check on the scaling laws found in
\cite{Niemeyer2} and to investigate the effect on them both of starting
the initial perturbations earlier, as pure growing modes within the linear
regime, and also of including vacuum energy (as a cosmological constant
term). For our calculations, we used a null-slicing code, modified from
that of Miller \& Motta (1989) \cite{Miller1}, which implements an
essentially identical method to that used in \cite{Niemeyer2} (but with
one key technical difference which we discuss in Section 2). This is not
as sophisticated in its treatment of the hydrodynamics as the Hawke \&
Stewart code \cite{Hawke} but is satisfactory for our purposes and has
some useful advantages particularly as regards the specification of
initial conditions. We have examined both super-critical and sub-critical
perturbations. The most notable results are: {\it (i)} we find good
agreement with the results of \cite{Niemeyer2} when we use the same
initial data; {\it (ii)} we obtain $\delta_c \simeq 0.45$ when we use
initial data specified as purely growing-mode perturbations, consistent
with the results of Green et al. \cite{Green3}; {\it (iii)} when a
cosmological constant term $\Lambda$ is included, $\delta_c$ increases
linearly with increasing $\Lambda$; {\it (iv)} subcritical collapse with
$\delta$ sufficiently close to $\delta_c$ produces strong compressions and
rarefactions before the overdensity subsides back into the surrounding
medium.

The organisation of the paper is as follows. In Section 2, we review the
mathematical formulation of the problem and the calculation method used;
Section 3 presents results from our calculations of PBH formation and
sub-critical collapse; Section 4 contains discussion and conclusions. We
use units for which $c = G = 1$ except where otherwise stated.


\section{Mathematical formulation \& calculation method}
\label{equations}
 For our calculations we have followed a procedure very similar to that of
Niemeyer \& Jedamzik \cite{Niemeyer2}. We will not repeat their full
discussion of the method but nevertheless it can be useful to make some
further comments about it here.

As with most of the other literature on this subject, we are restricting
attention to spherical symmetry, which very greatly simplifies the
calculations, and we have used the formulations of the relativistic
hydrodynamical equations given by Misner \& Sharp (1964) \cite{Misner} and
Hernandez \& Misner (1966) \cite{Hernandez}. Both of these are Lagrangian
formulations, the first using a diagonal metric (with the time referred to
as ``cosmic time'' which reduces to the familiar FRW time coordinate in
the absence of perturbations), and the second using an outward null
slicing where the time coordinate is an ``observer time'' (the clock time
as measured by a distant fundamental observer). The cosmic-time
formulation is particularly simple and has the advantage of using a
slicing which many people find intuitive; this was the approach used by
May \& White (1966) \cite{May} in their classic paper studying
spherically-symmetric gravitational collapse. However, as mentioned above,
this approach has a well-known drawback for studying black hole formation
in that singularities are typically formed rather early in calculations of
continuing collapse and it is not then possible to follow the subsequent
evolution. The outward null slicing approach is particularly convenient
for calculations involving black hole formation in spherical symmetry:
anything which could not be seen by a distant observer (e.g. singularity
formation) does not occur within the coordinate timespan, while all
observable behaviour can be calculated. This is, in some sense, the
optimal approach for studying black hole formation in spherical symmetry
as seen by an outside observer (being linked directly to potential
observations) although more sophisticated slicing conditions have
advantages for calculations away from spherical symmetry.

Following the introduction of the observer-time approach by Hernandez \&
Misner in 1966 \cite{Hernandez}, it was implemented soon afterwards in
unpublished calculations. A brief presentation of some results was given
by Miller \& Sciama (1980) \cite{Miller4} and a full discussion of the
technique used and of results obtained was given subsequently by Miller \&
Motta (1989) \cite{Miller1}. A problem with the use of this method
concerns the satisfactory specification of initial conditions which is not
natural to do on a null slice. Because of this, Miller \& Motta
\cite{Miller1} made a preliminary calculation using the Misner \& Sharp
\cite{Misner} formulation in order to construct data on an outgoing null
slice from initial conditions specified on a space-like slice; the
observer-time calculation then proceeded from the null-slice data
constructed in this way. Subsequently, Baumgarte, Shapiro \& Teukolsky
(1995) \cite{Baumgarte} made calculations using a similar technique and it
was these which were used as a reference point by Niemeyer \& Jedamzik
\cite{Niemeyer2}.

For the calculations of PBH formation reported in this paper, we have used
a modification of the Miller \& Motta code \cite{Miller1} (designed for
studying the collapse of an isolated object surrounded by vacuum) together
with elements from the codes by Miller \& Pantano (1990) \cite{Miller2}
and Miller \& Rezzolla (1995) \cite{Miller3} (designed for following
phase-transition bubble growth within an expanding universe).

In the remainder of this section, we give a brief overview of the
equations used and of our numerical code. The reader is referred to the
papers quoted above for further details. 

\subsection{The Misner-Sharp equations}
 For calculations in spherical symmetry, it is convenient to divide the
collapsing matter into a system of concentric spherical shells and to
label each shell with a Lagrangian co-moving radial coordinate which we
will denote with $r$. The metric can then be written in the form
 \begin{equation}
ds^2=-a^2\,dt^2+b^2\,dr^2+R^2\left(d\theta^2+\sin^2\theta 
d\varphi^2\right),
\end{equation}  
 where $R$ (the Schwarzschild circumference coordinate), $a$ and $b$ are
functions of $r$ and the time coordinate $t$. This was the form used 
Misner \& Sharp (1966) \cite{Misner}. 

For a classical fluid, composed of particles with nonzero rest-mass, it is
convenient to use the rest-mass $\mu$ contained interior to the surface of
a shell (or, equivalently, the baryon number) as its co-moving coordinate
$r$. For the case of a radiation fluid (as studied in this paper),
rest-mass and baryon number are not available as conserved quantities to
be used in this way but a similar procedure can still be followed by
introducing the concept of a conserved number of unit co-moving fluid
elements (Miller \& Pantano 1990) \cite{Miller2}. Denoting a ``relative
compression factor'' for these fluid elements by $\rho$ (equivalent to the
rest-mass density in the standard treatment), one then has
 \begin{equation}
d\mu=4\pi\rho R^2b\,dr, 
\end{equation}
 and identifying $\mu$ and $r$ then gives 
 \begin{equation}
b=\frac{1}{4\pi R^2\rho}.
\end{equation}
 Following the notation of \cite{Misner}, we write the equations in terms 
of the operators
 \begin{equation}
D_t\equiv\frac{1}{a}\left(\frac{\partial}{\partial
t}\right), \label{D_t}
\end{equation}
 \begin{equation}
D_r\equiv\frac{1}{b}\left(\frac{\partial}{\partial\mu}\right), \label{D_r}
\end{equation}
 and applying these to $R$ gives
 \begin{equation}
D_t R \equiv U, \label{U}
\end{equation}
 \begin{equation}
D_r R \equiv \Gamma, 
\label{gamma1}
\end{equation}
 where $U$ is the radial component of the four-velocity in the associated
Eulerian frame, using $R$ as the radial coordinate, and $\Gamma$ is a
generalisation of the Lorentz factor.

We are mostly dealing with processes occurring in the radiation dominated
era of the Universe for which the equation of state of the matter can be
written as
 \begin{equation}
p=\frac{1}{3}\,e,
\label{eq.state}
\end{equation}
 where $p$ is the pressure and $e$ is the energy density.  For
one-parameter equations of state of the form $p=p(e)$, the system of
Einstein and hydrodynamic equations can then be written as:
 \begin{equation}
D_tU=-\left[\frac{\Gamma}{(e+p)}D_rp+\frac{M}{R^2}+4\pi Rp\right], 
\label{Euler1}
\end{equation}
 \begin{equation}
D_t\rho=-\frac{\rho}{\Gamma R^2}D_r(R^2U),\label{D_trho}
\end{equation}
 \begin{equation}
D_t e=\frac{e+p}{\rho}D_t\rho,\label{D_te}
\end{equation}
 \begin{equation}
D_r a=-\frac{a}{e+p}D_r p,\label{D_ra}
\end{equation}
 \begin{equation}
D_r M=4\pi\Gamma eR^2, \label{D_rM}
\end{equation}
 where $M$ is a measure of the mass-energy contained inside radius $\mu$
and $\Gamma$ can be calculated either from (\ref{gamma1}) or from the 
constraint equation
 \begin{equation}
\Gamma^2=1+U^2-\frac{2M}{R} \label{Gamma}.
\end{equation}

\subsection{The Hernandez-Misner Equations}

Because of the problems mentioned earlier, Hernandez \& Misner
\cite{Hernandez} introduced the concept of ``observer time'', using as the
time coordinate the time at which an outgoing radial light ray emanating
from an event reaches a distant observer\footnote{We note that a somewhat
similar approach, but based on a double null foliation, has recently been
used by Harada et al. \cite{Harada} for studying some different aspects of
PBH formation.}. In the original formulation, this observer was placed at
future null infinity but for calculations in an expanding cosmological
background we use an FRW fundamental observer sufficiently far from the
perturbed region to be unaffected by the perturbation. Along an outgoing
radial null ray we have
 \begin{equation}
a\,dt = b\,dr,
\end{equation}
 and we define the observer time $u$ by
 \begin{equation}
f\,du = a\,dt - b\,dr,
\end{equation}
 with $f$ being an integrating factor which needs to be determined. In 
terms of this, the metric becomes
\begin{equation}
ds^2=-f^2\,du^2-2fb\,dr\,du+R^2\left(d\theta^2+\sin^2\theta 
d\varphi^2\right),
\label{nullmetric}
\end{equation}
 which is no longer diagonal. The operators equivalent to (\ref{D_t}) 
and (\ref{D_r}) are now
 \begin{equation}
D_t \equiv \frac{1}{f}\left(\frac{\partial}{\partial u}\right),
\end{equation}
 \begin{equation}
D_k \equiv \frac{1}{b}\left(\frac{\partial}{\partial r}\right) = 4\pi\rho
R^2\left(\frac{\partial}{\partial \mu}\right),
\end{equation}
 where $D_k$ is the radial derivative across the null slice and the 
corresponding derivative across the Misner-Sharp space-like slice is given 
by 
 \begin{equation}
D_r=D_k-D_t.
\end{equation}
 The observer-time equations replacing the cosmic-time ones (\ref{Euler1})  
-- (\ref{D_rM}) are then:
 \begin{eqnarray}
D_tU=-\frac{1} {1-c_s^2}\left[\frac{\Gamma}{(e+p)}D_kp + \frac{M}{R^2} 
+ 4\pi Rp \right. \nonumber\\
\hspace{5truecm} 
\left. +\, c_s^2\left(D_kU+\frac{2U\Gamma}{R}\right)\right], 
\label{Euler2}
\end{eqnarray}
 \begin{equation}
D_t\rho=\frac{\rho}{\Gamma}\left[D_tU-D_kU-\frac{2U\Gamma}{R}\right],
\end{equation}
 \begin{equation}
D_te=\left(\frac{e+p}{\rho}\right)\,D_t\rho,
\end{equation}
 \begin{equation}
D_kf=\frac{f}{\Gamma}\left(D_kU+\frac{M}{R^2}+4\pi Rp\right), 
\label{D_kf}
\end{equation}
 \begin{equation}
D_kM=4\pi R^2[e\Gamma-pU], \label{D_kM}
\end{equation}
 where $c_s = \sqrt{(\partial p/\partial e)}$ is the sound speed, which is
equal to $1/\sqrt{3}$ in the present case. The quantity $\Gamma$ is given
by equation (\ref{Gamma}), as before, and also by
 \begin{equation}
\Gamma=D_kR - U\,.
\end{equation}
 Using equations (24), (25) and (14), it is possible to derive the
following alternative equation for $f$:
 \begin{equation}
D_k\left[\frac{(\Gamma + U)}{f}\right] = -4\pi R(e+p)f\,.
\label{fnew}
\end{equation}
 In calculations concerning collapse of an isolated object surrounded by
vacuum in an asymptotically-flat spacetime, the observer time is taken to
be the clock time of a static observer at future null infinity and so
$(\Gamma + U)/f = 1$ at the location of that observer (since $\Gamma = 1$,
$U = 0$ and $f = 1$ there). It then follows from equation (\ref{fnew})
that $(\Gamma + U)/f = 1$ also at the surface of the collapsing object,
since the right hand side of (\ref{fnew}) is zero in vacuum. The condition
\begin{equation}
f = \Gamma + U,
\label{fbound}
\end{equation}
 at the surface is used as a boundary condition for $f$ and the values of
$f$ internal to that are then calculated from equation (\ref{D_kf}).

The general condition for a trapped surface is $D_k R \leq 0$. With
outgoing null slicing, $D_k R = 0$ should be reached only asymptotically
in the future, accompanied by the lapse $f$ going to zero, and $D_k R$
should never become negative. In practice, care is required in order to
achieve the exact synchronization of $D_k R \to 0$ and $f \to 0$ in a
numerical solution where the equations are discretized; if the
synchronization is not achieved, negative values of $D_k R$ do appear and
the evolution becomes unphysical.

In the case of an isolated collapsing object surrounded by vacuum, using
boundary condition (\ref{fbound}) together with equation (\ref{D_kf})  
ensures the correct behaviour. However, for the present situation, where
the surroundings are not vacuum and the spacetime is not asymptotically
flat, it is necessary to proceed in a different way. We are wanting to
synchronise the ``observer time'' with the clock time of a co-moving FRW
fundamental observer at the outer edge of the grid (setting $f = 1$ there)
and then to calculate $f$ elsewhere using this as a boundary condition.
For doing this, we found it essential to use equation (\ref{fnew}), which
guarantees synchronisation of $D_k R \to 0$ and $f \to 0$, rather than
equation (\ref{D_kf}), which always eventually gave rise to unphysical
behaviour with $D_k R < 0$. We think that this is a crucial point in using
an outward null slicing technique for any situation regarding black hole
formation within non-vacuum surroundings. It will apply equally to
calculations of black hole formation from core-collapse of high-mass
stars.

\subsection{The calculation method}

As mentioned above, our calculations of PBH formation have been made using
an explicit Lagrangian hydrodynamics code based on that of Miller \& Motta
(1989) \cite{Miller1} but with the grid organised in a way similar to that
in the code of Miller \& Rezzolla (1995) \cite{Miller4}, which was
designed for calculations in an expanding cosmological background. The
reader is referred to those papers for full details of the methods used
and we repeat just some main points here. The method described by Niemeyer
\& Jedamzik \cite{Niemeyer2}, following from the Baumgarte et al. code
\cite{Baumgarte}, is similar in most respects. In order to achieve good
grid coverage, we have used a composite prescription for the grid spacing,
with $\Delta \mu$ increasing exponentially going outwards through the
inner and outer parts of the grid but remaining constant in the
intermediate region. We start our perturbations in the linear regime with
length-scales larger than the cosmological horizon radius $R_H$, and have
the grid reaching out to $10\,R_H$ with around 2000 grid points.

For collapses leading to black hole formation, our calculations proceed in
two stages: first, initial data is specified on a space-like slice at
constant cosmic time, specifying the energy density $e$ and the
four-velocity component $U$ as functions of $R$ at an initial time $t_i$.  
This data is then evolved using the Misner-Sharp equations of Section 2.1
so as to generate a second set of initial data on a null slice (at
constant observer time). To do this, an outgoing radial light ray is
traced out from the centre and parameter values are noted as it passes the
boundary of each grid zone. The second set of initial data, constructed in
this way, is then evolved using the Hernandez-Misner equations of Section
2.2. For sub-critical cases, which end eventually with dispersal of the
perturbation into the surrounding uniform medium, we have continued with
the Misner-Sharp cosmic-time approach throughout.

The unperturbed background model is taken as a spatially-flat FRW model,
for which $\Gamma = 1$ (giving $U=\sqrt{2M/R}$) with $e(r) =$ constant at
any particular cosmic time $t$. The perturbations of $e$ and $U$ are then
superimposed on this background in the way described in the next section.
During the following evolutions, the metric functions $a$ (for cosmic
time) and $f$ (for observer time) are set equal to unity at the outer
boundary of the grid, thus synchronising the cosmic and observer times
with local clock time there as measured by the local co-moving FRW
fundamental observer.



\section{Description of the calculations}

In this section, we describe our calculations for both super-critical and
sub-critical perturbations.

\subsection{Initial Conditions}    
 As initial conditions for our calculations, we use spherically symmetric
density perturbations superimposed on a uniform FRW background. The
perturbations are specified in terms of the dimensionless quantity
 \begin{equation}
\delta_e(R) \equiv \frac{e - e_b}{e_b},
\label{delta_e}
\end{equation}
 where $e_b$ is the uniform background density given by $e_b = 3/32\pi t$
at any particular time $t$. Niemeyer \& Jedamzik \cite{Niemeyer2} used
three different types of expression for $\delta_e$ (shown in Figure 1 of
their paper):

\medskip\noindent
{\it Gaussian:}
 \begin{equation}
\delta_e(R)=A\exp\left(-\frac{2R^2}{(LR_H)^2}
\right),
\label{Gauss}
\end{equation}

\medskip\noindent
{\it Mexican hat:}
 \begin{equation}
\delta_e(R)=A\left(1-\frac{R^2}{L^2R_H^2}\right)
\exp\left(-\frac{3R^2}{2L^2R_H^2}\right),\label{Mexican}
\end{equation}

\medskip\noindent
{\it Polynomial:}
\begin{equation}
\delta_e(R)=\left\{\begin{array} {ll}
{\displaystyle{\frac{A}{9}\left(1-\frac{R^2}{L^2R_H^2}\right) 
\left(3-\frac{R^2}{L^2R_H^2}\right)}}
 & \hbox{\rm if \ } R<\sqrt{3}\,(LR_H)\nonumber\\ \\
0 & \hbox{\rm if \ } R\ge\sqrt{3}\,(LR_H)
\end{array} \right.
\label{Polynomial}
\end{equation}
 where the parameters $A$ and $L$ set the amplitude and length-scale of
the perturbation. For the Mexican-hat and polynomial perturbations, the
excess energy in the overdense region is exactly balanced by the deficit
in the outer underdense region (i.e. $\int_0^{\infty} 4\pi \delta_e R^2dR
= 0$), whereas the Gaussian ones have only an excess, decreasing
asymptotically to the background value. The latter is not very
satisfactory for cosmological perturbations and so we concentrate here on
the first two types.

We follow the previous literature in also using an ``integrated''
perturbation $\delta$ which represents the mass-energy excess in the
overdense region with respect to that in a corresponding uniform solution:
 \begin{equation}
\delta\equiv\frac{\displaystyle{\int_0^{R_o} 4\pi
\delta_e R^2dR}}{\displaystyle{\frac{4}{3}\pi
R_o^3}},
\label{delta}
\end{equation}
 where $R_o$ is the radius of the overdensity. (In the case of the
Gaussian, $R_o$ is defined as the radius at which $\delta_e$ has fallen to
$1/e^2$ of its value at $R = 0$.)

We are mostly starting with perturbations which are still on length-scales
much larger than the horizon scale [$R_o/R_H \ge 5$] and are well within
the linear regime [typically $\delta \sim 10^{-2}$]. Setting initial
conditions within the linear regime makes it easy to specify consistent
density and velocity perturbations representing a purely growing mode:
$\delta_e$ grows linearly with time and the associated velocity
perturbation is given roughly by
 \begin{equation}
\delta_u(R) = - \frac{\delta_e}{4},
\label{delta_u1}
\end{equation}
 (coming from formulae in \cite{Padmanabhan}), where $\delta_u$ is defined
in the same way as for $\delta_e$ in equation (\ref{delta_e}):
 \begin{equation}
\delta_u(R) \equiv \frac{U - U_b}{U_b},
\label{delta_u2}
\end{equation}
 with $U_b(R)=HR$ being the velocity field of the background Hubble flow.
(We are here considering adiabatic perturbations with a Newtonian gauge,
in which the perturbations do not generate off-diagonal components of the
metric.) The corresponding decaying modes have decreasing amplitudes and,
if excited by some process, would disappear again rather rapidly. Of the
perturbations produced from inflation, only growing modes are of interest
at the times which we are considering here.

\subsection{Evolution of supercritical perturbations}

In this sub-section, we describe results from representative evolutions
leading to black hole formation, carried out using the null-slicing
formulation.

Before moving on to new calculations, we first needed to check on whether
our code did reproduce the results of Niemeyer \& Jedamzik
\cite{Niemeyer2} when we used their choice of initial conditions and a
simple exponential grid similar to theirs. We found extremely close
agreement. For perturbations with the initial $\delta$ only slightly
larger than the critical value $\delta_c$, the masses of the black holes
produced, $M_{BH}$, follow a scaling law:
 \begin{equation}
M_{BH}=K(\delta-\delta_c)^{\gamma}\, M_H(t_H),
\label{scaling-law}
\end{equation}
 where $K$ and $\gamma$ are constants and $M_H(t_H)$ is the cosmological
horizon mass at the horizon-crossing time $t_H$ (i.e. when $R_o(t_H) =
R_H(t_H)$). The type of behaviour given by (\ref{scaling-law}) is familiar
from the literature on critical collapse (which, however, is generally
considering collapse under simpler circumstances and not within the
context of an expanding universe). Our results match very closely with
those of \cite{Niemeyer2}, with $\delta_c \simeq 0.67$ for the Mexican-hat
profile and $\simeq 0.71$ for the polynomial and Gaussian profiles and
with $\gamma$ between $0.36 - 0.37$ in each case. These values for
$\gamma$ are very close to the value $0.356...$ calculated
semi-analytically for the same equation of state within the standard
critical collapse scenario \cite{Koike}.

\begin{figure}[ht!]
\centerline{\psfig{file=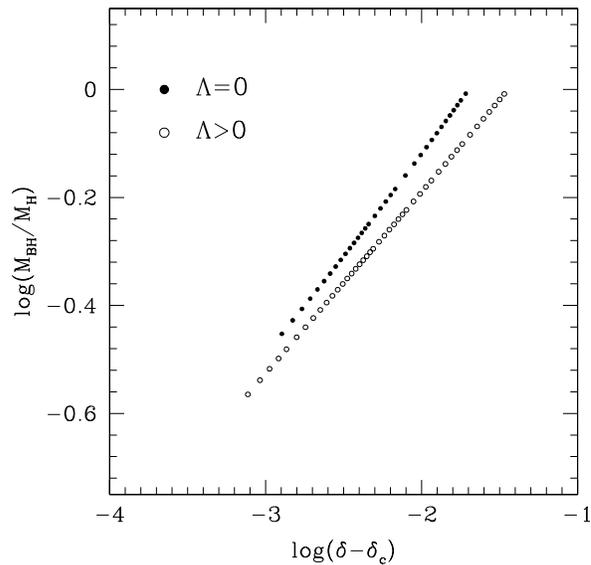,width=8cm}}
\caption{\label{Fig.1}\small 
 Scaling behaviour for $M_{BH}$ as a function of $(\delta-\delta_c)$
calculated for growing-mode Mexican-hat perturbations specified within the
linear regime. The filled circles refer to the standard calculation
discussed in section 3.2, while the open circles are for a calculation
including a non-zero cosmological constant $\Lambda$, as discussed in
section 3.3, giving $y = 3.0 \times 10^{-3}$.
 }
 \end{figure}

Following these initial calculations to reproduce previous results, we
then carried out further ones starting from growing-mode perturbations
specified within the linear regime and with length-scales larger than the
cosmological horizon. These perturbations were evolved with the
Misner-Sharp code until the moment when they entered the horizon and the
current value of $\delta$ was then calculated for use as our measure of
perturbation amplitude in the discussion of scaling laws. Having done
this, we switched to the Hernandez-Misner code for completing the
calculation.

Plotting the eventual black hole mass against $(\delta - \delta_c)$, we
obtained scaling curves very similar to those of \cite{Niemeyer2} (and
with almost identical values of $\gamma$) but with substantially different
values for $\delta_c$: for Mexican-hat perturbations, we found
$\delta_c\simeq 0.43$ and for polynomial perturbations, $\delta_c\simeq
0.47$. Our scaling-law results for Mexican-hat perturbations are shown as
the $\Lambda = 0$ curve in Figure \ref{Fig.1}. The reason for the changed
values of $\delta_c$ is clear: in \cite{Niemeyer2} the initial
perturbations, specified at the horizon-crossing time, had part of their
amplitude contributed by a decaying-mode component which then rapidly
decreased leaving only the growing-mode part visible. Only the part of the
perturbation amplitude corresponding to the growing mode is relevant for
the black hole formation and so the effective $\delta$ is smaller than
that calculated in \cite{Niemeyer2}.

Noting the work of Hawke \& Stewart (2002) \cite{Hawke} and the fact that
the horizon-scale is an important length-scale in the problem, we do not
expect that the linear scaling law will continue to indefinitely small
values of $M_{BH}$ and $(\delta - \delta_c)$ but, instead, that it will
level off at some minimum value of $M_{BH}$. Confirming this behaviour in
the case of growing-mode perturbations starting in the linear regime is of
great interest but would require refinement of our present code to
increase resolution and improve the capacity for handling strong shocks.

\begin{figure}[hp!]
\centerline{\psfig{file=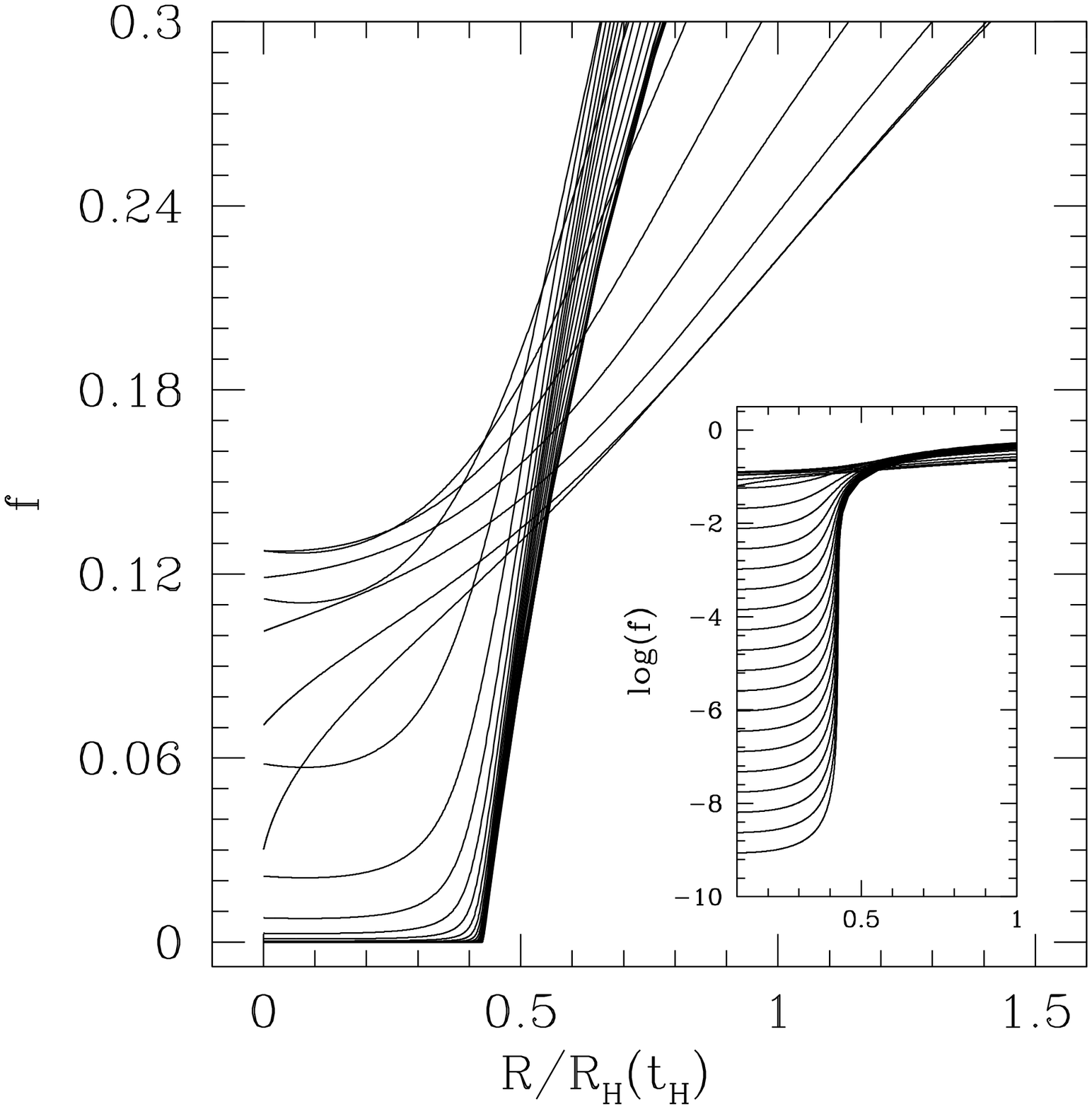,width=6.5cm} \ \
	    \psfig{file=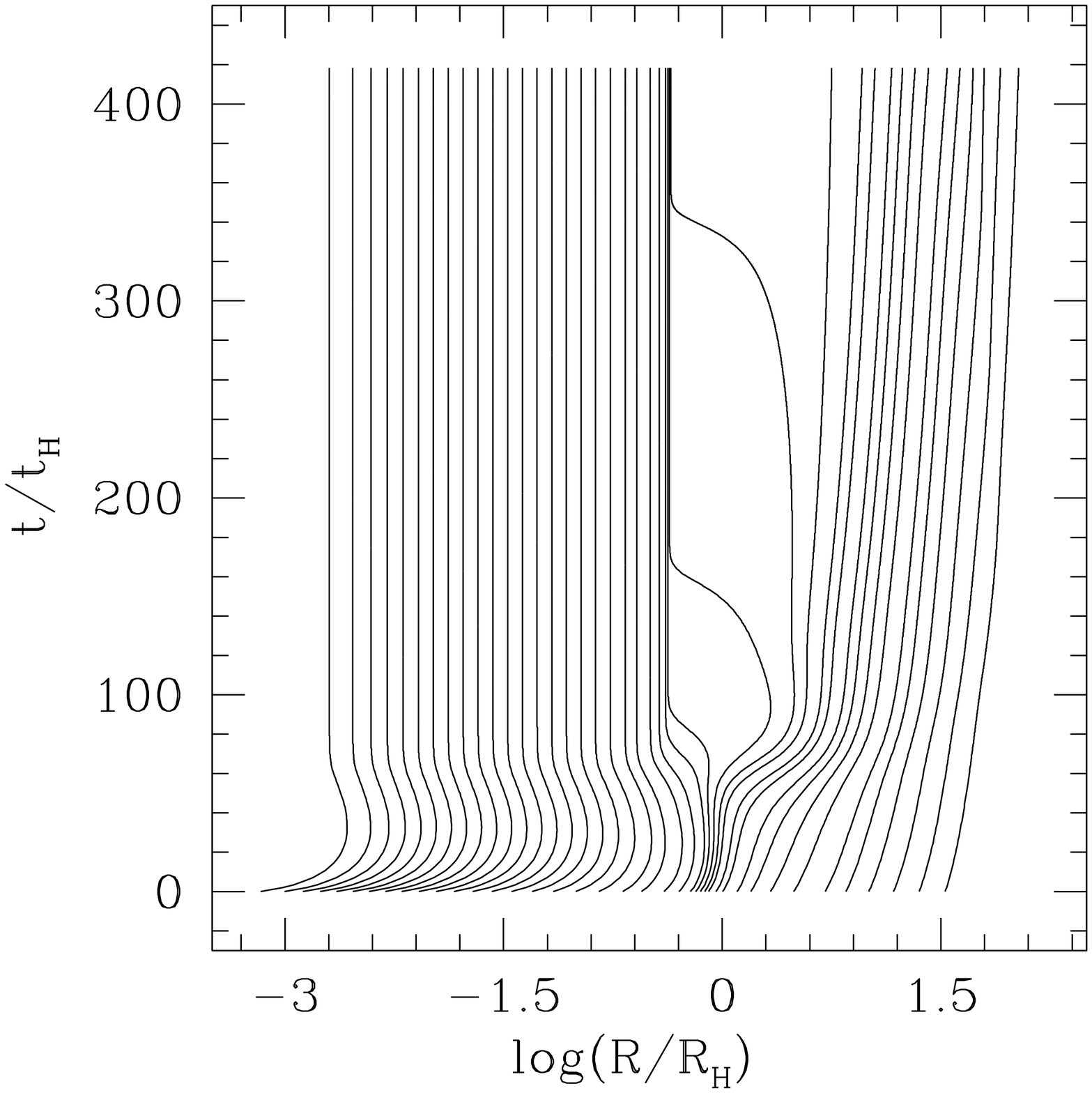,width=6.5cm}}
\centerline{\psfig{file=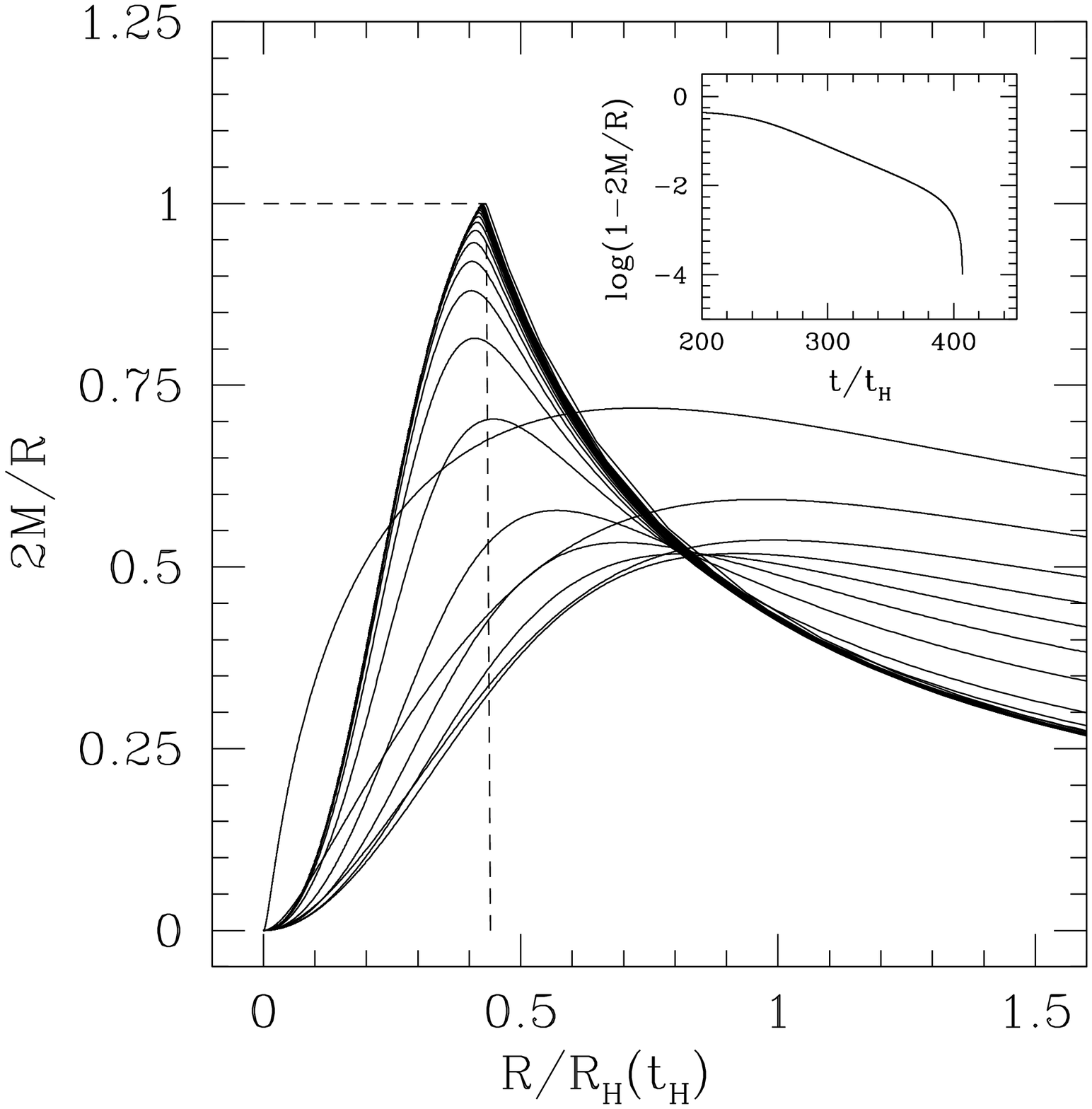,width=6.5cm} \ \
	    \psfig{file=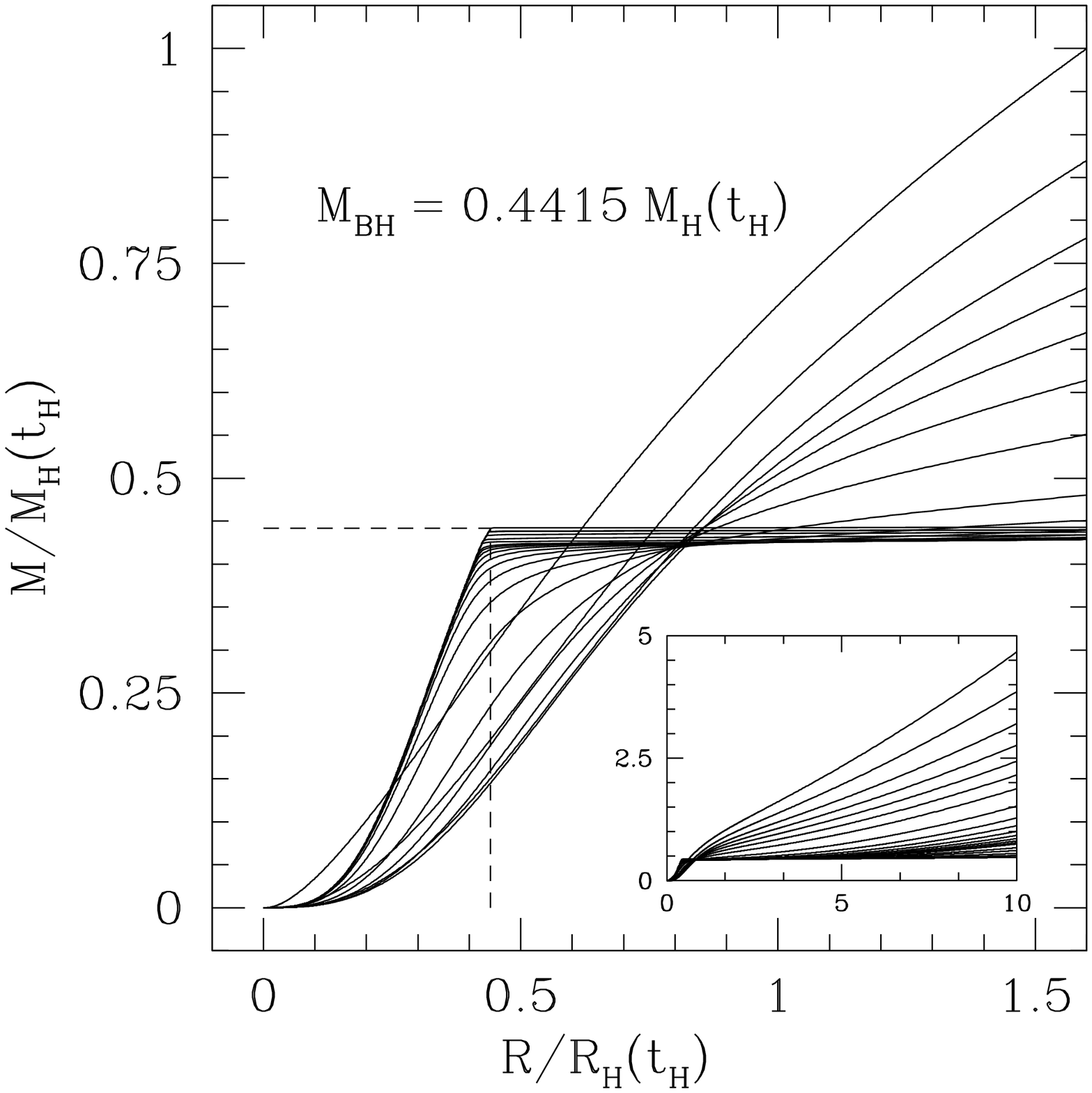,width=6.5cm}}

\caption{\label{Fig.2}\small 
 A typical evolution leading to black hole formation: the initial
perturbation had a Mexican-hat profile and gave $(\delta-\delta_c)=2.37
\times 10^{-3}$ at the horizon crossing time. The top left-hand panel
shows the behaviour of the lapse function (the time sequence of the curves
goes from bottom to top on the right hand side); the top right-hand panel
shows the fluid-element worldlines (the time is measured in units of the
horizon-crossing time $t_H$). The bottom left-hand panel shows the profile
of $2M/R$ at different times, with the inset showing the approach of the
maximum value of $2M/R \to 1$; the bottom right-hand panel shows the
corresponding evolution of the mass-energy (in both of these panels, the
time sequence of the curves goes from top to bottom on the right hand
side).
 }

\end{figure}

Figure \ref{Fig.2} shows some more detailed results for a particular
representative case within the linear scaling regime. This run starts from
a growing-mode Mexican-hat perturbation with $R_o/R_H=5$ giving
$(\delta-\delta_c)= 2.37\times 10^{-3}$ at horizon crossing, which leads
to formation of a black hole with $M_{BH} = 0.4415\, M_H(t_H)$. The top
two panels show the evolution of the lapse $f$ and the corresponding
behaviour of the fluid worldlines. The interpretation of the collapse of
the lapse is particularly clear when an observer-time formulation is used:
as $f \to 0$, the redshift of outgoing signals increases and the evolution
as seen by a distant observer becomes frozen, corresponding to black hole
formation (see also the small inset where $\log f$ is plotted). Note,
however, that strictly ``black hole formation'' occurs only asymptotically
in the future according to this formulation. In the plot for the
worldlines, one can see the separation between the matter which goes to
form the black hole and the matter which continues to expand, with a
semi-evacuated region being formed between them. Note that some of the
outer material first decelerates but then accelerates again before
crossing this semi-evacuated region to fall onto the black hole. The
bottom left-hand panel of Fig.~\ref{Fig.2} shows the behaviour of the
ratio $2M/R$, plotted against $R$ at successive times. The event horizon
corresponds to the asymptotic location of the outermost trapped surface,
where $D_k R = 0$ and $R = 2M$. For the present purposes, we need an
operational definition for calculating $M_{BH}$, bearing in mind that the
black hole is only formed asymptotically and that further material may
continue to accrete. The inset in the bottom left-hand panel of
Fig.~\ref{Fig.2} shows the approach of the maximum value of $2M/R \to
1$: our operational definition for $M_{BH}$ is to set it equal to the
value of $M$ at the maximum of $2M/R$ when $(1-2M/R)$ there first becomes
smaller than $10^{-4}$. The bottom right-hand panel shows a corresponding
plot for $M$ against $R$. It can be seen that the profiles for $M$ become
very flat just outside the black hole region at late times, a consequence
of the very low densities being reached there (less than $10^{-4}$ of the
background density at the horizon-crossing time). The small inset shows
the continuation of this figure up to larger radial scales. For
calculations with $\delta$ closer to $\delta_c$, the rarefactions formed
become increasingly deep, and one sees strong shock waves appearing at the
outer edge of the under-dense region.

\subsection{Evolution of super-critical perturbations when $\Lambda > 0$}

We were also interested to investigate the effect for PBH formation of
including a cosmological constant large enough to affect the dynamics. We
recognise that this is a highly idealised scenario since the present-day
cosmological constant would have had negligible effect in the early
universe and other vacuum energies present after inflation are unlikely to
have been constant in time (e.g. quintessence). However, it is of
conceptual interest to find out what the behaviour would be in this
hypothetical case. A cosmological constant $\Lambda$ is equivalent to a
false vacuum with energy density $e_v = \Lambda/8\pi$ and pressure $p_v =
- \Lambda/8\pi$. Its effect can be included by adding these terms onto the
standard energy density $e$ and pressure $p$ wherever those appear.

A positive $\Lambda$ eventually causes the expansion of the universe as a
whole to start accelerating and acts against the growth of overdensities,
while a negative $\Lambda$ would aid general collapse. (The equations
governing the background expansion when $\Lambda \neq 0$ are summarised in
the Appendix.) As the background density decreases with time, the
cosmological constant becomes progressively more important until, when
$e_v$ becomes greater than $e$, the deceleration is reversed and becomes
an acceleration [see equation (\ref{eq.Fried2+})]. It is convenient to
introduce a quantity $y$, the ratio between the vacuum energy and the
total energy in the uniform background
\begin{equation}
y \equiv {{e_v}\over{e + e_v}},
\end{equation}
 which can then be written as
\begin{equation}
y = \frac{4}{3}\Lambda M_H^2, 
\end{equation}
 since $M_H = {4\over 3}\pi {R_H}^3 (e + e_v)$ with $R_H = 2M_H$ (Note
that this type of relation holds for the cosmological horizon in the same
way as for a black hole event horizon)\footnote{Both are trapped surfaces.
A black hole event horizon is the asymptotic location of the outermost
trapped surface for outgoing light-rays whereas the cosmological horizon
is the innermost trapped surface for incoming light rays.}. In the
following, we will use $y$ as a general measure of the importance of the
$\Lambda$ term, with $M_H$ being measured at the horizon-crossing time for
a perturbation with $\delta = \delta_c$. The influence of $\Lambda$ during
formation of a black hole of mass $M_{BH}$ can similarly be characterised
by the quantity $\Lambda M_{BH}^2$ and so, for a given $\Lambda$, is
greatest for large black holes.
 
In making computations with $\Lambda > 0$, it is particularly important to
start the calculation at an early time when the perturbation has a
length-scale larger than the horizon. For appreciating its effects, one
wants $\Lambda$ to be sufficiently large so as to make a significant
difference for the collapse, but not so large that it creates a problem
for constructing the null-slice initial data. By starting the calculation
sufficiently early, this can be achieved although the values of $y$ which 
we will be considering are all very small (in the range $10^{-3} -   
10^{-2}$).

The qualitative picture of collapses leading to black hole formation is
not changed very greatly by the presence of a $\Lambda$ term but there are
significant differences in the parameters of the scaling law. We started
all of these calculations with perturbations at five times the horizon
scale (i.e. $R_o/R_H = 5$).

The impact of $\Lambda$ on the scaling law can be seen in Figure
\ref{Fig.1}: $\gamma$ decreases for $\Lambda > 0$ and for sufficiently
small $\Lambda$ is found to follow a linear relationship
 \begin{equation}
\gamma(\Lambda) \simeq \gamma(0) - 8.3\,y .
\end{equation}
 The critical amplitude $\delta_c$ {\em increases} with increasing
$\Lambda$ and also follows a linear relationship:
 \begin{equation}
\delta_c(\Lambda) \simeq \delta_c(0) + 0.98\,y .
\end{equation}
 This behaviour can be interpreted as follows: a positive $\Lambda$ acts
against collapse, so that corresponding black hole masses will be lower
and the threshold amplitude $\delta_c$ will be raised. For a given
$\Lambda$, its influence is greater for larger black-hole masses than for
smaller ones ($\propto M_{BH}^2$) and this gives rise to the observed
decrease in $\gamma$.



\subsection{Evolution of subcritical perturbations}

For subcritical perturbations with $\delta$ considerably less than
$\delta_c$, the perturbation initially grows but then subsides back into
the surrounding medium in an uneventful way. However, for perturbations
with $\delta$ sufficiently close to $\delta_c$, some very interesting
behaviour is seen and we present results from a representative case of
this in the present subsection. Our calculations for subcritical 
perturbations use only the Misner-Sharp code.

\begin{figure}[ht!]
\centerline{\psfig{file=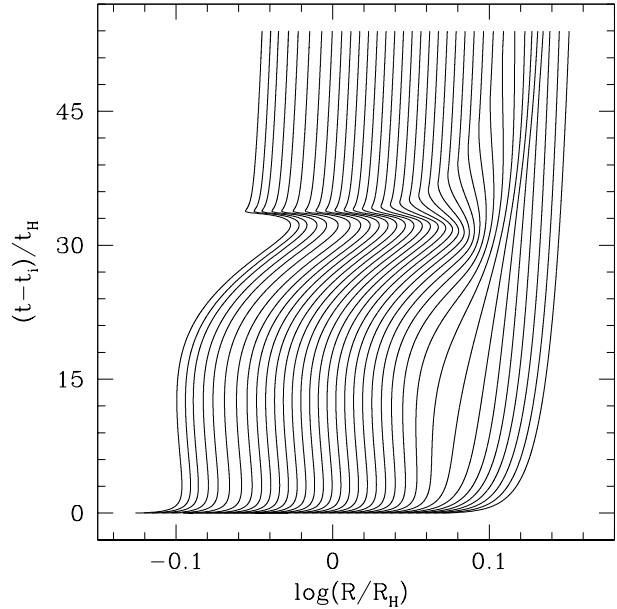,width=8.5cm}}
\caption{\label{Fig.3}\small 
 Worldlines for a Mexican-hat perturbation with $(\delta - \delta_c)=-3.0
\times 10^{-3}$). This plot shows alternating collapse and expansion of
the perturbed region while the outer material continues to expand
uniformly. The ``cosmic'' time is measured in units of the time at horizon
crossing.
 }
 \end{figure}

\begin{figure}[hp!]
\vskip -1.0cm
\centerline{\psfig{file=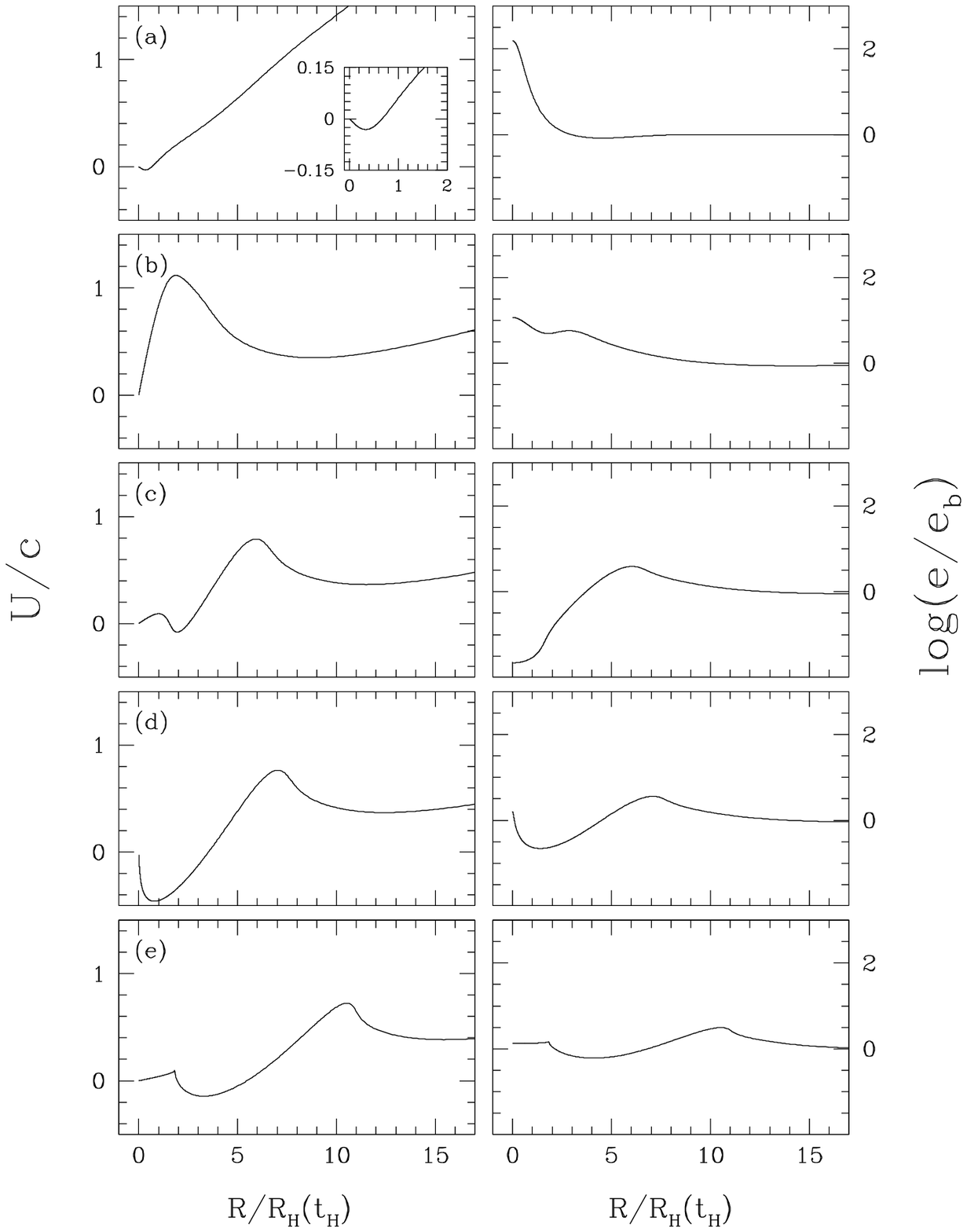,width=22.cm}}
\vskip -1.0cm
\caption{\label{Fig.4} \small 
 Plots of local quantities as functions of $R/R_H(t_H)$: the velocity
$U/c$ is shown in the left-hand column and the energy density $e/e_b$ in
the right-hand column. The frames correspond to the following values of
$(t-t_0)/t_H$: (a) 7.02; (b) 25.92, (c) 31.67; (d) 33.64; (e) 40.11\,.
Note that $R_H$ is increasing with time and so points with $R/R_H(t_H) >
1$ can be within the current horizon scale at times after
horizon-crossing.
}
 \end{figure}

\begin{figure}[ht!]
\centerline{\psfig{file=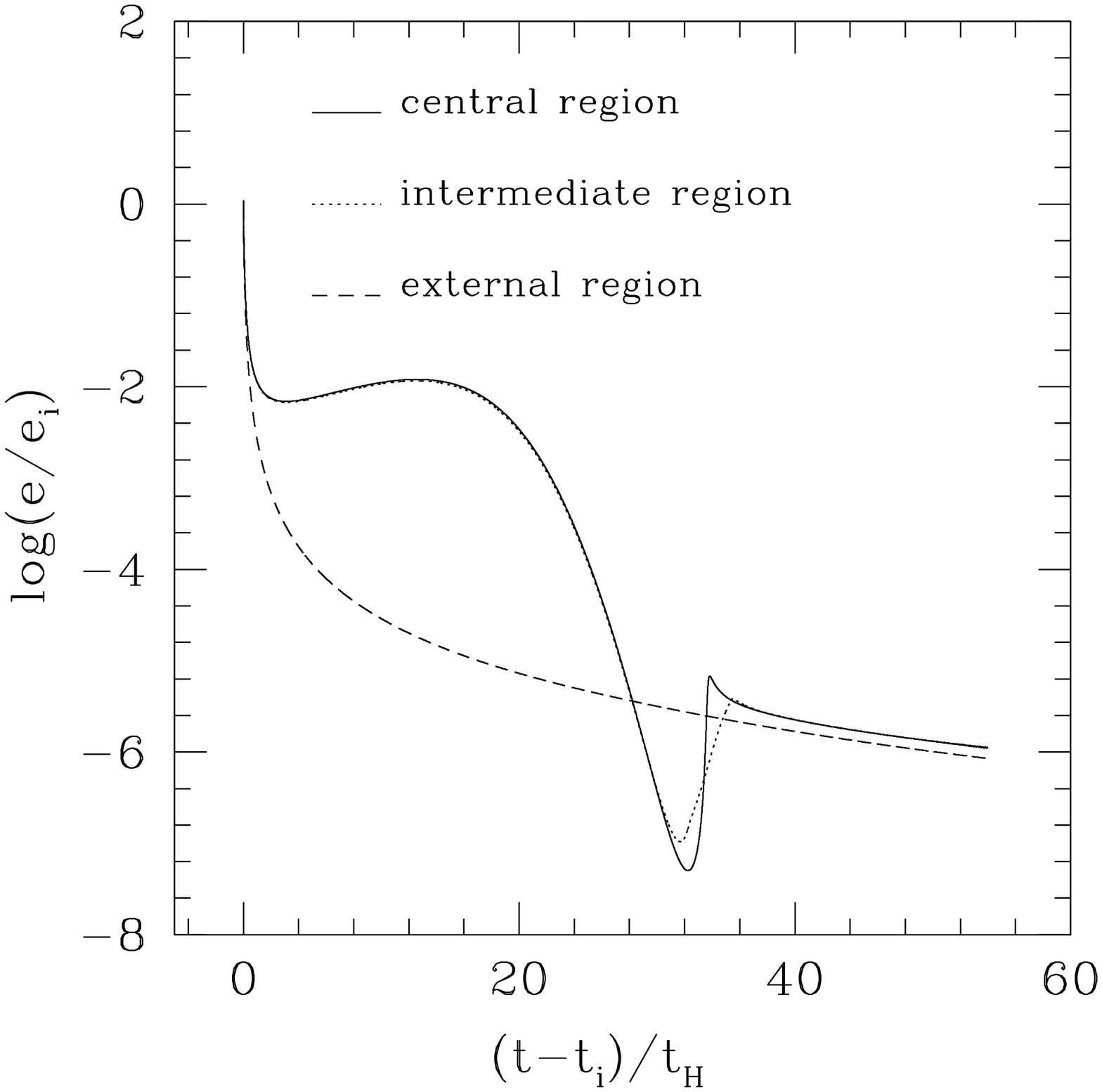,width=6.5cm} \ \
            \psfig{file=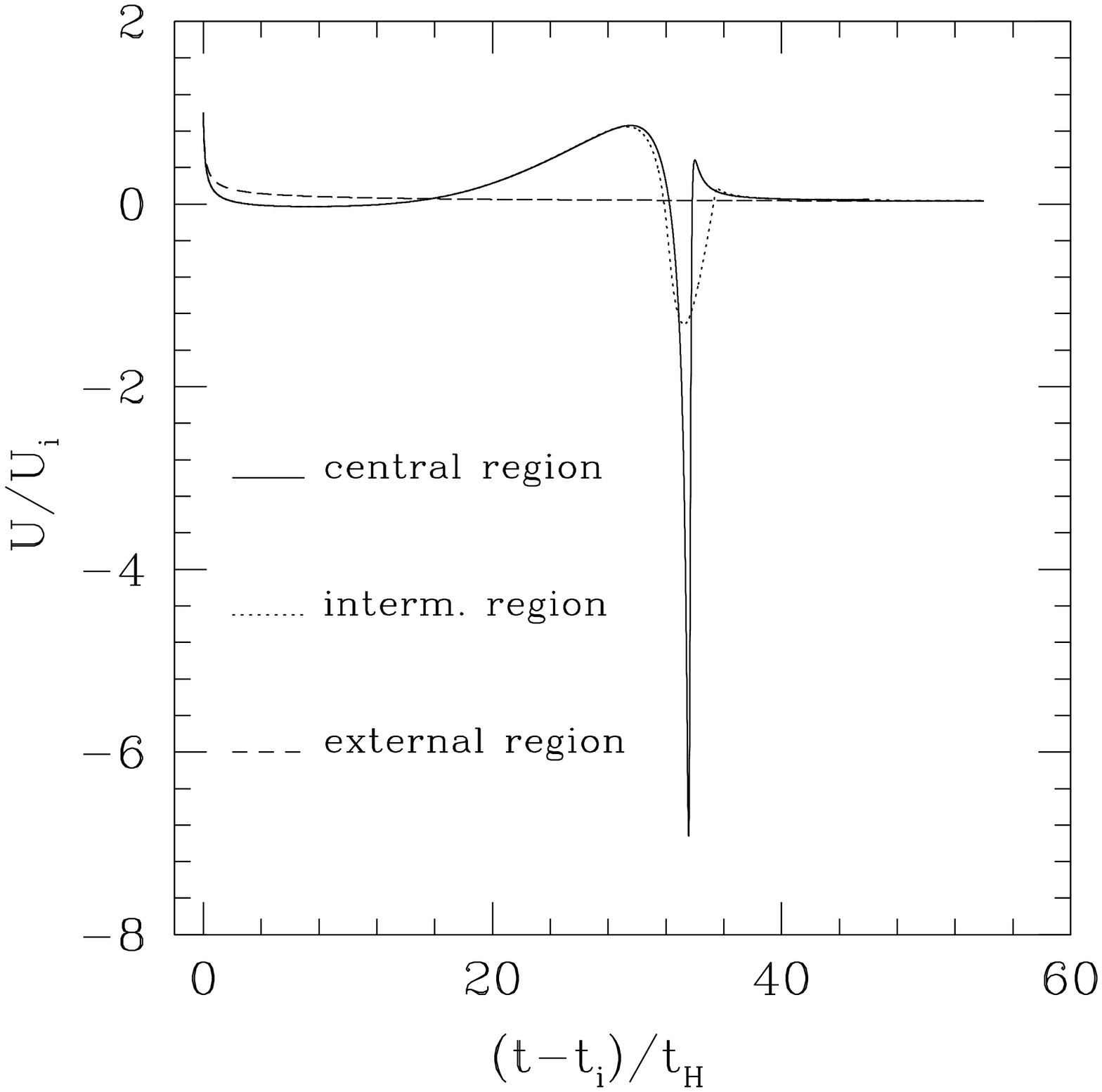,width=6.5cm}}
\caption{\label{Fig.5} \small
 Evolution of the energy density $e$ and radial velocity $U$ at three
(comoving) locations: near to the centre of the perturbation, at an
intermediate region and at the edge of the grid where the fluid is
unperturbed. Each quantity is measured in units of its initial value at
the same comoving location.
}
 \end{figure}

The run presented starts with a Mexican-hat perturbation specified in the
linear regime, with ($\delta - \delta_c = -3 \times 10^{-3}$). In Figure
\ref{Fig.3}, the fluid worldlines are plotted and the main features of
interest can already be seen from this. Figures \ref{Fig.4} and
\ref{Fig.5} then show details of the evolution of the energy density $e$
and the radial velocity $U$. Figure \ref{Fig.4} shows a sequence of
snap-shots of these quantities, plotted as a functions of $R$, at key
moments during the evolution. Figure \ref{Fig.5} shows the time evolution
of these quantities at three (comoving) locations: near the centre of the
perturbation, at an intermediate region (mid-way through the collapsing
matter) and at the edge of the grid where the fluid is unperturbed.

The evolution can be summarized in terms of the following steps (see 
particularly Figures \ref{Fig.3} and \ref{Fig.4}): 

\begin{enumerate}

\item
 Initially, the perturbation has very small amplitude ($\delta_e \sim
10^{-2}$) and its length-scale is five times the horizon scale; the
perturbation amplitude then grows within the expanding fluid. The
deceleration in the perturbed region is larger than that in the
unperturbed region and its expansion lags progressively behind that of the
outer matter until eventually it starts to re-contract shortly after
horizon crossing. The maximum infall velocity reached is, however, rather
small [see row (a) of Figure \ref{Fig.4} which is for a time considerably
after horizon crossing, when the perturbation has become very nonlinear].
The infall can be clearly seen in Figures \ref{Fig.3} and \ref{Fig.4} but
is only just visible in the second frame of Figure \ref{Fig.5}.

\item 
 The contraction is not strong enough to produce a black hole and the
fluid bounces out again [row (b)], expanding until it encounters the
surrounding matter which did not participate in the contraction. A
compression wave forms where the two regions of fluid meet, while the
density becomes very low at the centre of the perturbation [row (c)].

\item  
 The compression wave proceeds out into the surrounding material [row (d)]
but also some matter is sent back into the middle of the rarefaction where
it undergoes a second bounce which is much more extreme than the first
with a very abrupt change of velocity in the central regions (as can be
seen from Figure \ref{Fig.5}). Whereas the outward moving compression is
damped geometrically as it proceeds to spherical surfaces with
progressively larger areas, the inward-moving wave of material is
geometrically amplified by the inverse process. The reason for the second
collapse and bounce being more violent than the first is that while the
first is a collapse of an overdensity which is resisted throughout by
internal pressure, the second is essentially the collapse of a ``shell''
with near vacuum inside it and is hence close to free-fall until just
before the bounce.

\item 
 The compression wave formed by the second bounce propagates out into the
surrounding medium following the first one [row (e)]. Both proceed to
damp geometrically and eventually the medium returns to a uniform state.
Note that the second compression wave is very steep fronted (see Figure
\ref{Fig.4}) but is not quite a shock. It is likely that that genuine
shocks would be seen for perturbations with $\delta$ closer to $\delta_c$.

 \end{enumerate}    

It is useful to make some further comments about the plots in Figure
\ref{Fig.5}. The left-hand plot shows the density normalized to its
initial value at the same comoving location: this shows the variation of
the local density against the background of the general decrease in
density as the universe expands and allows one to see clearly the local
contraction and expansion of the fluid. Initially, the perturbed region is
also still expanding but it then starts to contract when the perturbation
amplitude has reached a high enough value. The right-hand plot shows the
local value of the radial velocity also normalized to its initial value at
the same co-moving location. This has a very different behaviour from that
of the energy density, with the inward velocity during the second collapse
being very much greater than that during the first and the second bounce
being far more abrupt. We see just two bounces in our calculation: after
this the expansion of the universe prevents further ones, although we
expect that more bounces would be seen for an initial perturbation with
$\delta$ closer to $\delta_c$.



\section{Conclusions}

We have presented results from calculations of PBH formation in an
expanding universe during the radiation dominated era, following a broadly
similar approach to that used by Niemeyer \& Jedamzik \cite{Niemeyer2} in
previous investigations of this subject but with a number of key
differences. Growing-mode perturbations were introduced into an otherwise
uniform expanding medium and their subsequent behaviour was then followed.
Those overdensities with amplitudes $\delta$ greater than the critical
value $\delta_c$ give rise to black hole formation whereas smaller ones
contract to a maximum compression and then bounce, eventually dispersing
into the surrounding medium. Defining $\delta$ to be the fractional mass
excess within the overdensity at the time of horizon crossing, we find
that $\delta_c$ ranges between $0.43$ and $0.47$, depending on the
perturbation shape. These values are consistent with the results of
Shibata \& Sasaki \cite{Shibata}, obtained with a different method,
resolving a previously noted discrepancy \cite{Green3}. We have verified
the results of \cite{Niemeyer2} concerning the scaling law relation
between the black-hole mass $M_{BH}$ and $(\delta - \delta_c)$ when the
latter is sufficiently small. The power of the scaling law $\gamma$ is
essentially unchanged for our modified initial conditions, in contrast
with the large changes seen for $\delta_c$. We note, however, the
important results of Hawke \& Stewart (2002) \cite{Hawke} who found that,
for a class of perturbations specified in a non-linear regime at
sub-horizon scales, the scaling law does not continue to arbitrarily small
$M_{BH}$ but levels off at around $10^{-4}$ of the horizon mass due to
interaction with the surrounding medium. Our code does not yet have the
capability of approaching sufficiently close to $\delta_c$ to reveal this
effect; it would be important for future work to establish to what extent
the Hawke \& Stewart \cite{Hawke} result is dependent on the form of the
initial conditions.

We have investigated the effect on the scaling law of introducing a
cosmological constant term $\Lambda$ and found that positive values of
$\Lambda$ give rise to lower values of $\gamma$ and higher values of
$\delta_c$, each varying linearly with $\Lambda$ when the latter is
sufficiently small. This can be understood in terms of the effect of
a cosmological constant in opposing collapse.

We also studied sub-critical collapses (with $\delta < \delta_c$). When
$\delta$ is close to $\delta_c$, we find that this can be a surprisingly
violent process. The initial collapse and bounce are rather mild, being
moderated by the internal pressure, but after the subsequent reflection
from the surrounding medium and re-collapse (with matter falling into
near-vacuum), the second bounce can be very violent with a rapid velocity
change and the sending out of a second compression wave into the
surrounding medium following the one produced by the first bounce. After
this, the possibility of further bounces is stopped by the overall
background expansion for the cases studied (although additional bounces
are expected for $\delta$ closer to $\delta_c$).

There are a number of topics concerning PBH formation which may be of
great interest for cosmology and which we are intending to study further,
including: determination of the proportion of the matter in the universe
going into black holes according to various scenarios for the
perturbations coming from inflation (with the possibility of ruling out
some scenarios); investigation of possible enhanced PBH formation at the
time of phase transitions (see Jedamzik \& Niemeyer \cite{Jedamzik});
alternative possibilities for the formation of intermediate mass black
holes or of the seeds for super-massive black holes at the centres of
galaxies (as mentioned by Bicknell \& Henriksen \cite{Bicknell} in 1979).
Work is now in progress to investigate these.

\vspace{0.3cm} \noindent 
 {\bf Acknowledgements:} In the course of this work, we have benefited
from helpful discussions with many colleagues including Bernard Carr,
Alexander Polnarev, Marco Bruni, Karsten Jedamzik, Ian Hawke and Carlo
Baccigalupi.

\appendix
\section{Cosmological solution with non-zero $\Lambda$}

 We present here some of the analytic expressions describing an expanding
universe with $\Lambda\neq0$ in the radiation-dominated era. These
equations are certainly not being presented here for the first time but
they are not easy to find in the usual cosmology textbooks and we think
that it may be useful to present them together here. In this part only, we
use physical units and do not set $c = G = 1$.

First, we note the forms taken by the Friedman equation and the associated 
acceleration equation (we are taking the spatially flat case):
 \begin{equation}
\left(\frac{\dot{a}}{a}\right)=\frac{8\pi
G}{3c^2}e+\frac{\Lambda c^2}{3},
\label{eq.Fried+}
\end{equation} 
 \begin{equation}                                           
\left(\frac{\ddot{a}}{a}\right)=-\frac{8\pi G}{3c^2}
\left(e - \frac{\Lambda c^4}{8\pi G} \right),
\label{eq.Fried2+}
\end{equation}  
where $e$ is the energy density of radiation which scales as
\begin{equation}
e=e_i\left(\frac{a_i}{a}\right)^4,
\end{equation}
(here the subscript $i$ refers to a fiducial initial time).
Inserting this into (\ref{eq.Fried+}), we obtain the integral equation
 \begin{equation}
\displaystyle{\int_0^{(a_i/a)}}\frac{\displaystyle{\frac{a_i}{a^\prime}\,
d\left(\frac{a_i}{a^\prime}\right)}}{\displaystyle{\sqrt{\frac{8\pi
G}{3c^2}e_i+\frac{\Lambda
c^2}{3}\left(\frac{a_i}{a^\prime}\right)^4}}}=\int_0^tdt^\prime.
\label{inteq}
\end{equation}
 If $\Lambda>0$, the solution for the scale factor is 
\begin{equation}
a(t)=a_i\left(\frac{8\pi Ge_i}{\Lambda c^4}\right)^{1/4}
\left[\sinh\left(2\sqrt{\frac{\Lambda}{3}}ct\right)\right]^{1/2},
\label{a-eds+}
\end{equation}
Using (\ref{a-eds+}) in (\ref{eq.Fried+}) we get the Hubble parameter
\begin{equation}
H(t)=\sqrt{\frac{\Lambda}{3}}c\coth\left(2\sqrt{\frac{\Lambda}{3}}ct\right),
\label{H-eds+}
\end{equation}
which, in the limit $\Lambda \to 0$, reduces to the standard
expression $H(t)=1/2t$. Inserting (\ref{H-eds+}) into
(\ref{eq.Fried+}) we get the 
expression for $e(t)$
\begin{equation} 
e(t)=\frac{\Lambda c^4}{8\pi G}
\left[\sinh\left(2\sqrt{\frac{\Lambda}{3}}ct\right)\right]^{-2}.
\label{e-eds+}
\end{equation}
Another useful expression is the inverse of (\ref{H-eds+})
\begin{equation}
t=\left(4\sqrt{\frac{\Lambda}{3}}c\right)^{-1}
\ln\left(\frac{H+\sqrt{\frac{\Lambda}{3}}c}{H-\sqrt{\frac{\Lambda}{3}}c}
\right).,
\label{t-eds+}
\end{equation}
 which we have used in the cosmic time code to calculate the initial time
for the calculation.

Finally, for completeness, we calculate the expression for the two
cosmological horizons. From (\ref{H-eds+}) we get a straightforward the
expression for the Hubble horizon $R_H\equiv c/H$,
 \begin{equation}
R_H(t)=\left(\frac{\Lambda}{3}\right)^{-\frac{1}{2}}
\tanh\left(2\sqrt{\frac{\Lambda}{3}}ct\right)   
\label{R_H-eds+}.
\end{equation}
For the particle horizon, the calculation is more complicated. From the 
definition 
\begin{equation}
R_h\left(t\right) \equiv a\left(t\right)
\int_{0}^{t}\frac{cdt^\prime}{a\left(t^\prime\right)},
\label{R_h}
\end{equation}
we get
\begin{equation}
R_h\left(t\right)=
\left[\sinh\left(2\sqrt{\frac{\Lambda}{3}}ct\right)\right]^{1/2}
\int_{0}^{t}
\frac{du}{\left[\sinh\left(2\sqrt{\frac{\Lambda}{3}}cu\right)\right]^{1/2}}\,, 
\end{equation}
and, with the aid of integral tables, we get the final form of the 
solution
\begin{equation}
\displaystyle{R_h(t)=\left(2\sqrt{\frac{\Lambda}{3}}\right)^{-1}
\left[\sinh\left(2\sqrt{\frac{\Lambda}{3}}ct\right)\right]^{1/2}F(\phi,k)}
\end{equation}
where $F(\phi,k)$ is an incomplete elliptic integral of the first type: 
\begin{equation}
F(\phi,k)\equiv\int_0^\phi\frac{d\theta}{(1-k^2\sin^2\theta)^{1/2}} ,
\end{equation}
 with
\begin{equation}
\phi=\arccos\left[\frac{1-
\sinh\left(2\sqrt{\frac{\Lambda}{3}}ct\right)}{1+
\sinh\left(2\sqrt{\frac{\Lambda}{3}}ct\right)}\right] ,
\end{equation} 
 and
\begin{equation}
k=\frac{1}{\sqrt{2}}.
\end{equation} 

\noindent
 When $\Lambda<0$, relations (\ref{a-eds+}), (\ref{H-eds+}) and
(\ref{e-eds+}) are unchanged apart from replacing the hyperbolic functions
by the corresponding trigonometric ones. This is consistent with the
oscillating behaviour of a universe with $\Lambda<0$, characterized by a
sequence of expanding and contracting phases.



\section*{References}

\begin{thebibliography}{99}

\bibitem{Liddle2} Liddle A.R. \& Lyth D.H. 2000
\emph{Cosmological Inflation and Large-Scale Structure}
Cambridge University Press

\bibitem{Zeldovich} Zel'dovich Ya.B. \& Novikov I.D. 1966
\emph{Astron.Zh.} {\bf{43}} 758 [\emph{Sov.Astron.}
{\bf{10}} 602 (1967)]    

\bibitem{Hawking} Hawking S.W. 1971
\emph{MNRAS} {\bf{152}} 75 

\bibitem{Carr1} Carr B.J. \& Hawking S.W. 1974 \emph{MNRAS}
{\bf{168}}, 399 

\bibitem{Nagatani} Nagatani Y. 1999 \emph{Phys.Rev.D} {\bf{59}} 041301

\bibitem{Cline1} Cline D.B., Sanders D.A. \& Hong W.P. 1997
\emph{Astrophys. J.} {\bf{486}} 169 

\bibitem{Cline2} Cline D.B., Matthey C. \& Otwinowsky S. 1999
\emph{Astrophys.J.} {\bf{527}} 827

\bibitem{Green2} Green A.M. 2002 \emph{Phys.Rev.D} {\bf 65} 027301

\bibitem{Carr4} Carr B.J. 2003 \emph{Lect. Notes Phys.} {\bf
631} 301

\bibitem{Barrau} Barrau A., Blais D., Boudoul G. \&
Polarski D. 2003 \emph{Phys. Lett. B} {\bf 551} 218

\bibitem{McGibbon} MacGibbon J.H. 1987 \emph{Nature} {\bf
329} 308 

\bibitem{Blais} Blais D., Kiefer C. \& Polarski D. 2002
\emph{Phys. Lett. B} {\bf 535} 11

\bibitem{Afshordi} Anshordi N., McDonald P., Spergel
D.N. 2003 \emph{Astrophys. J.} {\bf 594} L71
	
\bibitem{Carr2} Carr B.J. 1975 \emph{Astrophys.J.} {\bf{201}} 1

\bibitem{Kim1} Kim H.I. \& Lee C.H. 1996 \emph{Phys.Rev.D} {\bf{54}} 6001

\bibitem{Nakamura} Nakamura T., Sasaki M., Tanaka T. \& Thorne
K.S. 1997 \emph{Astrophys.J.Lett.} {\bf{487}} L139

\bibitem{MACHO} MACHO Collaboration, Alcock C. et al. 1996
\emph{Astrophys.J.} {\bf{471}} 774

\bibitem{Green4} Green A.M. \& Jedamzik K. 2002 \emph{Astron.Astrophys.} 
{\bf 395} 31

\bibitem{Rahvar} Rahvar S. 2003 \emph{Int.J.Mod.Phys.D} {\bf 12} 45  

\bibitem{Carr3} Carr B.J., Gilbert J.H. \& Lidsey J.E. 1994
\emph{Phys.Rev.D} {\bf{50}} 4853

\bibitem{Green1} Green A.M. \& Liddle A.R. 1997 \emph{Phys.Rev.D}
{\bf{56}} 6166

\bibitem{Liddle1} Liddle A.R. \& Green A.M. 1998 \emph{Phys.Rep.}
{\bf 307} 125

\bibitem{Kribs} Kribs G.D., Leibovitch A.K. \& Rothstein I.Z. 1999 
\emph{Phys.Rev.D} {\bf{60}} 103510 

\bibitem{Bringmann} Bringmann T., Kiefer C. \& Polarski D. 2002
\emph{Phys.Rev.D} {\bf 65} 024008

\bibitem{Nadezhin} Nadezhin D.K., Novikov I.D. \& Polnarev A. G. 1978 
\emph{Astron.Zh.} {\bf{55}} 216 [\emph{Sov.Astron.} {\bf{22(2)}} 129 
(1978)]

\bibitem{May} May M.M. \& White R.H. 1966 \emph{Phys.Rev.} {\bf{141}} 1232

\bibitem{Podurets} Pondurets M.A. 1964 \emph{Astron.Zh.} {\bf{41}} 1090
[\emph{Sov.Astron.} {\bf{8}} 868 (1965)]
 
\bibitem{Bicknell} Bicknell G.V. \& Henriksen R. N. 1979
\emph{Astrophys.J.} {\bf{232}} 670

\bibitem{Niemeyer1} Niemeyer J.C. \& Jedamzik K. 1998
\emph{Phys.Rev.Lett.} {\bf{80}} 5481

\bibitem{Niemeyer2} Niemeyer J.C. \& Jedamzik K. 1999 \emph{Phys.Rev.D}
{\bf{59}} 124013

\bibitem{Choptuik} Choptuik M.W. 1993 {Phys.Rev.Lett.} {\bf{70}} 9

\bibitem{Evans} Evans C.R. \& Coleman J.S. 1994 {Phys.Rev.Lett.} 
{\bf{72}} 1782

\bibitem{Gundlach} Gundlach C. 1999 \emph{Living Rev.Rel.} 
[{\tt{http://www.livingreviews.org/lrr-1999-4}}]

\bibitem{Hernandez} Hernandez W.C. \& Misner C.W. 1966 \emph{Astrophys.J.}
{\bf{143}} 452

\bibitem{Harada} Harada T., Goymer C. \& Carr B.J. 2002
\emph{Phys. Rev. D} {\bf 66} 104023 \\
Harada T. \& Carr B.J. 2004 astro-ph/0412134 \\
Harada T. \& Carr B.J. 2004 astro-ph/0412135

\bibitem{Hawke} Hawke I. \& Stewart J. M. 2002 \emph{Class. Quantum
Grav.} {\bf{19}} 3687
 
\bibitem{Shibata} Shibata M. \& Sasaki M. 1999 \emph{Phys.Rev.D}
{\bf{60}} 084002

\bibitem{Green3} Green A.M., Liddle A.R., Malik K.A., Sasaki M. 2004 
\emph{Phys.Rev.D}, {\bf 70}, 041502 

\bibitem{Miller1} Miller J.C. \& Motta S. 1989 \emph{Class. Quantum
Grav.} {\bf{6}} 185

\bibitem{Misner} Misner C.W. \& Sharp D.H. 1964 \emph{Phys.Rev} {\bf{136}} B571

\bibitem{Miller4} Miller J.C. \& Sciama D.W. 1980 in
\emph{General Relativity and  
Gravitation: one hundred years after the birth of Albert Einstein - Vol.2} 
ed. Held A. (Plenum Press: New York), p. 359

\bibitem{Baumgarte} Baumgarte T.W., Shapiro S.L. \& Teukolsky S.A. 1995
\emph{Astrophys.J.} {\bf{443}} 717

\bibitem{Miller2} Miller J.C. \& Pantano O. 1990 \emph{Phys.Rev.D}
{\bf{42}} 3334

\bibitem{Miller3} Miller J.C. \& Rezzolla L. 1995 \emph{Phys.Rev.D}
{\bf{51}} 4017
 
\bibitem{Padmanabhan} Padmanabhan T. 1993 \emph{Structure Formation in
the Universe} Cambridge University Press

\bibitem{Koike} Koike T, Hara T. \& Adachi S. 1995 \emph{Phys.Rev.Lett.}
{\bf{74}} 5170

\bibitem{Jedamzik} Jedamzik K. \& Niemeyer J.C. 1999 \emph{Phys.Rev.D}
{\bf{59}} 124014

\end{thebibliography}
\end{document}